\documentstyle[11pt,newpasp,twoside]{article}
\begin{document}
\title{The TexOx Survey of Radio-Selected Galaxy Clusters}
 \author{Steve Croft, Steve Rawlings}
\affil{Astrophysics, Physics Department, Keble Road, Oxford OX1 3RH, UK}
\author{Gary J.\ Hill, Pamela L.\ Gay, Joseph R.\ Tufts}
\affil{McDonald Observatory, Department of Astronomy,
University of Texas at Austin, RLM 15.308, TX 78712-1083, USA}

\begin{abstract}
We present some initial results from the TexOx
(Texas--Oxford) Cluster (TOC) survey -- a new
method of selecting distant galaxy clusters.
The cosmic evolution of the radio source population suggests that
some massive clusters at high redshift will contain several radio-loud
AGN. We searched for extreme over-densities at $\sim$ mJy
levels in $7\arcmin \times 7\arcmin$ boxes within the
NVSS radio catalogue, covering a large ($\sim 1100 ~ \rm deg^2$) sky area.
We have acquired optical images for $\sim$130 cluster candidates,
and followed up a subset of these with the VLA, and
with Calar Alto near-IR imaging.
Ryle Telescope observations have yielded at least one Sunyaev-Zel'dovich (SZ)
detection of a massive $z \sim 1$ system. Spectroscopic
follow-up with 8-m class telescopes is in progress.

\end{abstract}

\section{Introduction}

Traditional methods of finding clusters, such as searching for peaks in
galaxy counts within wide-field optical surveys, become increasingly
difficult at high redshift due to contamination by foreground
galaxies. The redshift distribution of faint radio AGN means that
contamination effects are much reduced for any search for
an over-density at redshift $z \sim 1$. The availability of wide-field
radio surveys like the $1.4 ~ \rm GHz$ NRAO VLA Sky
Survey (NVSS; Condon et al.\ 1998), suggests an alternative way
of finding clusters which we have been pursuing.

\section{The TexOx Cluster Survey and Follow-up Observations}

If the NVSS catalogue were gridded into $7\arcmin \times 7\arcmin$ boxes, we
would expect the mean number of sources per box to be $\sim 0.8$. We created
a catalogue of candidate clusters by searching within $7\arcmin \times
7\arcmin$ boxes, centred on each NVSS source, for an additional four or
more sources. In some cases, large low-redshift radio structures are
seen by NVSS as several ``individual'' sources,
but these were easily eliminated from our cluster sample by
inspection of the NVSS images. The remainder of the fields were
our candidate galaxy clusters. At the depth of the NVSS, there is a
reasonable probability that the richest clusters will contain several
radio-loud AGN, and this probability is not a strong function of redshift out
to $z \sim 1$ because of the cosmic evolution of the radio source population.

The cluster candidates were imaged over 30 clear dark nights with
the 2.7-m telescope at McDonald Observatory, with typical exposures of
$30 - 60$ minutes
per band in $R$ and $I$ (reaching a limiting sensitivity of $R \sim 24$).
We recover some known low-redshift clusters, including several
in the Abell catalogue at $z \sim 0.1$, and the $z=0.37$ cluster around
the quasar 3C48. Many of the fields contained optically-obvious clusters
near the magnitude limit of our observations. Although any survey using
AGN to find clusters will find only a subsample of the total
population, the technique has proved very efficient at finding real clusters.

The cluster radio sources are typically associated with galaxies of
$R \sim 23$, with the brightest cluster member a magnitude or two
brighter, suggesting a typical redshift around $0.5$. Importantly, about
$25\%$ of TexOx clusters have no radio source identifications to the
limits of our 2.7-m observations, suggesting they are at $z > 0.7$.
For those fields which we were convinced
contained real but distant clusters, we used the Calar Alto 3.5-m telescope
to obtain near-IR images. Our collaborators in Cambridge then
used the Ryle telescope to try to detect Sunyaev-Zel'dovich
(Sunyaev and Zel'dovich, 1972) decrements in some of these. 
This yielded a detection of an object
with a minimum gas mass of around $5 \times 10^{13} M_{\sun}$, in which
five of the NVSS sources are identified with host galaxies of
$R \approx 24, J \approx 20, K \approx 18$ --
presumably a rich cluster at $z \sim 1$ (Cotter et al.\ 2001; Croft et al.\
2001 in prep.). We are
using Gemini and the 9.2-m Hobby-Eberly Telescope to get spectroscopic
redshifts for this and similar systems.

\section{Conclusions}

Targeting over-densities in the NVSS provides a powerful way to
find high-redshift clusters, with different biases to
X-ray, optical or infrared approaches.
Our TexOx Cluster (TOC) survey will provide a unique
sample of clusters, including some at $z > 1$, which can be used to
confirm predictions of Large Scale Structure simulations,
to analyse cluster properties, and to study the correlation between
cluster dynamics, the intra-cluster gas, and AGN activity.

\acknowledgments

We thank Niv Drory, our collaborator on the Calar Alto programme, and
Garret Cotter, Helen Buttery, Rhiju Das, Keith Grainge, William Grainger,
Michael Jones, Guy Pooley and Richard Saunders, our collaborators
in the Ryle Telescope group.
This material is based in part upon work supported by the Texas Advanced
Research Program under Grant No. 009658-0710-1999.
SC thanks PPARC for the support of a PhD studentship.

\end{document}